\documentstyle[prd,aps,preprint,tighten,floats]{revtex}
\begin{document}
\draft
\preprint{UPR-646-T}
\date{February 1995}
\title {KALUZA-KLEIN BLACK HOLES WITHIN HETEROTIC
STRING THEORY ON A TORUS}
\author {Mirjam Cveti\v c
\thanks{E-mail address: cvetic@cvetic.hep.upenn.edu}
and Donam Youm\thanks{E-mail address: youm@cvetic.hep.upenn.edu}}
\address {Physics Department \\
          University of Pennsylvania, Philadelphia PA 19104-6396}
\maketitle
\begin{abstract}
{We point out that in heterotic string theory compactified on a 6-torus,
after a consistent truncation of the 10-d gauge fields and the antisymmetric
tensor fields, 4-dimensional black holes of Kaluza-Klein theory on a 6-torus
constitute a subset of solutions.}
\end{abstract}

In this note we show that  4-dimensional (4-d) black holes
(BH) of Kaluza-Klein (KK) theory constitute a subset \cite{FIV} -
\cite{NONEX}  of 4-d BH solutions of an effective heterotic
string theory compactified on a 6-torus \cite{SEN}.

An effective 4-d action for the massless bosonic sector of
heterotic string vacua compactified on a 6-d torus is obtained
\cite{HETE} by compactifying the (massless) bosonic part of $D=10$ $N=1$
supergravity coupled to $N=1$ super Yang-Mills theory, containing
the dilaton $\Phi^{(10)}$, the 2-form field
$B_{\hat \mu \hat \nu}^{(10)}$ and 16 Abelian gauge fields
$A_{\hat \mu}^{(10)\,I}$ ($I=1,...,16$), on a 6-torus
\footnote{For notational conventions and the relationship of
the 4-d massless modes to the bosonic modes of the corresponding
10-d $N=1$ supergravity theory see for example Ref. \cite{SENII}.}:
\begin{equation}
S= \int {\rm d}^4 x \sqrt{-G} {\rm e}^{-\Phi}
[{\cal R}_G + \partial_{\mu} \Phi \partial^{\mu} \Phi
-{1 \over {12}} H_{\mu\nu\rho}H^{\mu\nu\rho}
- F^a_{\mu\nu}(LML)_{ab}(F^b)^{\mu\nu}
+ {1\over 8}{\rm Tr}(\partial_{\mu} ML \partial^{\mu}ML)]\ ,
\label{string}
\end{equation}
where $F^a_{\mu\nu} = \partial_{\mu} A^a_{\nu} - \partial_{\nu} A^a_{\mu}$
and $H_{\mu\nu\rho} =(\partial_{\mu} B_{\nu\rho} + 2A^a_{\mu} L_{ab}
F^b_{\nu\rho}) + permutations$ ($a, b = 1,...,28$).  $G \equiv {\rm det}
G_{\mu \nu}$ and the Ricci scalar ${\cal R}_G$ are defined in terms of the
string metric $G_{ \mu  \nu}$.  $M$ is the $O(6,22)$ matrix of
the following 28 scalar fields: the internal part of the 10-d
metric $\hat G_{mn}\equiv
G^{(10)}_{m+3, n+3}$ $(m, n = 1,...,6)$, ``antisymmetric'' background fields
$B_{mn}\equiv
B^{(10)}_{m+3,n+3}$ ($m,n=1,...,6$) and ``gauge'' background fields
$A_m^I\equiv A^{(10) \, I}_{m+3}$ ($m=1,...,6,\ \ I=1,...,16$).
$M$ has the properties:
\begin{equation}
MLM^T = L, \ \ \ \  M^T = M, \ \ \ \  L = \left (\matrix{0 & I_6 & 0 \cr
I_6 & 0 & 0 \cr 0 & 0 & -I_{16}} \right ) \  ,
\label{const}
\end{equation}
where $I_n$ is the $n \times n$ identity matrix.  $L$ is the matrix invariant
under $O(6,22)$ transformations.  The 4-d dilaton field
$\Phi \equiv \Phi^{(10)} -{1 \over 2}\ln {\rm det} \hat{G}$ is
defined in terms of the 10-d dilaton field $\Phi^{(10)}$ and determinant
of the internal metric $\hat G_{mn}$.  Gauge fields $A_\mu^m\equiv {1\over 2}
\hat G^{mn}G_{n+3,\mu}^{(10)}$ ($m,n=1,...,6$) are related to the
off-diagonal components of the 10-d metric.  Gauge fields $A_\mu^a$ with
$a=7,...,28$ are related to the off-diagonal components of the
10-d anti-symmetric tensor $B^{(10)}_{\hat \mu \hat \nu}$ and the 4-d
space-time components of the 10-d gauge fields $A_{\hat \mu}^I$.

We choose to set the 10-d Abelian gauge fields and 10-d 2-form fields
equal to zero; this choice is consistent with the equations of motion
in the  corresponding 10-d supergravity theory, and thus with the equations of
motion of the  4-d effective action (\ref{string}).
Consequently, a consistent truncation of (\ref{string})
corresponds to setting the anti-symmetric tensor field
$B_{\mu\nu}$, a set of 4-d gauge fields $A^a_{\mu}$ (a=7,...,28),
as well as the scalar background fields $B_{m n}$,
and $A_m^I$ to zero.  The action (\ref{string}) then reduces to the following
form: \begin{equation}
S =\int {\rm d}^4 x \sqrt{-G} {\rm e}^{-\Phi} [{\cal R}_{G} +
\partial_{\mu} \Phi \partial^{\mu}\Phi - F^a_{\mu\nu}(LML)_{ab}
(F^b)^{\mu\nu}+{1\over 8}{\rm Tr}(\partial_{\nu} ML
\partial^{\mu} ML)] \ ,
\label{kstring}
\end{equation}
where
\begin{equation}
M = \left ( \matrix{\hat{G}^{-1} & 0 & 0 \cr 0 & \hat{G} & 0
\cr 0 & 0 & I_{16}} \right ) \ .
\label{moduli}
\end{equation}
depends only on $\hat G_{mn}$, a real symmetric $(6\times 6)$-matrix
of scalar fields associated with the internal metric of 6-tori.
The action (\ref{kstring}) can now be written explicitly as:
\begin{eqnarray}
S &=  \int {\rm d}^4 x \sqrt{-G}{\rm e}^{-\Phi}\left [ {\cal R}_G
+\partial_{\mu}\Phi \partial^{\mu}\Phi -{1\over 4} \hat{G}_{mn}
F^m_{\mu\nu} F^{n\mu\nu} + {1\over 4} \partial_{\mu}\hat{G}_{mn}
\partial^{\mu} \hat{G}^{mn} \right ]\ \ \ \ \ \ \ \ \
\nonumber \\
&= \int {\rm d}^4 x \sqrt{-g} \left [ {\cal R}_g - {1\over 2}\partial_\mu
\tilde{\varphi} \partial^{\mu} \tilde{\varphi}
-{1\over 2}\partial_\mu \varphi \partial^{\mu} \varphi
-{1 \over 4}{\rm e}^{\alpha\varphi} \rho_{mn} F^m_{\mu\nu} F^{n\mu\nu}
+ {1\over 4}\partial_{\mu} \rho_{mn} \partial^{\mu} \rho^{mn}\right ] \ ,
\label{kk}
\end{eqnarray}
where $\rho_{mn}$ is the unimodular part of the metric $\hat G_{mn}$,
$\varphi={1\over\alpha}({1\over n}\ln\det\hat{G} - \Phi)$, and
$\tilde{\varphi}\equiv{1\over \alpha }(\sqrt{1\over {2n}}\ln\det\hat{G} +
\sqrt{2 \over n}\Phi)$.
Here, $\rho^{mn} \rho_{n\ell} = \delta^m_\ell$ and
$\alpha =\sqrt{{n+2}\over n }$ with $n=6$.
The scalar curvature ${\cal R}_g$ and $g = {\rm det}g_{\mu\nu}$ are
expressed in terms of the Einstein-frame metric $g_{\mu\nu}$.

The action (\ref{kk}) is that of 11-d KK theory
compactified on a 7-torus, where the gauge field $A_{\mu}^{7}$
associated with the seventh torus is turned off.
Consequently, the field $\tilde{\varphi}=\sqrt{2\over{n+2}}\Phi^{(10)}$,
parameterizing the size of the seventh torus, decouples (except for
4-d gravity) from the other fields and can therefore be set to a constant.
This result is obvious, once one realizes that the bosonic sector of 10-d
$N=1$ supergravity with the 10-d gauge fields and  antisymmetric
tensor field turned off corresponds to 11-d KK theory compactified
down to 10-d with the 10-d gauge field, associated with the
compactified dimension, turned off.

Thus, the action (\ref{kk}) is effectively that of 10-d KK theory compactified
on a 6-torus \cite{CHB}.  The corresponding 4-d BH solutions of
(\ref{kk}) are then those of ($4+n$)-d ($n$=6) KK theory.
In particular, with further consistent truncations of the gauge fields,
{\it i.e.}, $A_\mu^m=0$ ($m=1,...,k(<6)$), (\ref{kk}) reduces
to the effective action of ($10-k$)-d KK theory.  Namely, the corresponding
internal metric fields, {\it i.e.}, combinations of $\varphi$
and $\rho_{mn}$ ($m\ {\rm or}\ n \in \{1,...,k\}$), decouple from
the other fields (except for 4-d gravity) and can thus be set to constant
values. Specifically, for the choice of $k=5$ (only one non-zero
gauge field) (\ref{kk}) reduces to the action of an effective
5-d KK theory with the corresponding 4-d KK BH solutions \cite{FIV}, as
discovered by  Duff {\it et al.} \cite{DUFF}.

The supersymmetric embedding of the bosonic action (\ref{kk})
allows one to derive the Bogomol'nyi bound for the ADM mass of
the above class of spherically symmetric BH solutions. Among
them the supersymmetric ones, {\it i.e.}, those which preserve
(constrained) supersymmetry, can be regarded as nontrivial vacuum
configurations, since they saturate the corresponding Bogomol'nyi bounds.
Supersymmetric embedding \footnote{The embedding is a generalization of a
supersymmetric embedding for the BH solutions in $5$-d KK theory, found
by Gibbons and Perry \cite{GIBB}.} of 4-d Abelian KK BH's with diagonal
internal metric Ansatz has been carried out \cite{EX} within $(4+n)$-d
KK theory ($1\le n\le 11$), and can thus be applied to BH solutions of
(\ref{kk}) as well.  Such supersymmetric BH's have at most one
magnetic ($P$) and one electric ($Q$) charge arising from
different $U(1)$'s, thus corresponding to solutions
in the effective 6-d KK theory with the internal isometry
$U(1)_M \times U(1)_E$.  Embedding of (\ref{kk}) in 10-d $N=1$
supergravity ensures \cite{EX} that the resulting vacuum configuration
preserves one ($N=1$) of $N=4$ supersymmetries of the effective 4-d action.

The corresponding non-extreme solutions with a diagonal internal metric
\cite{WARSAW,NONEX} as well as a class of  those
with a non-diagonal internal metric \cite{NONEX} have also been found.
The latter ones can be obtained \cite{NONEX} as solutions of (\ref{kk}),
by performing $SO(n)$ ($n=6$) rotations on the solutions  with a
diagonal internal metric.  This $SO(6)$ symmetry is realized as
a subset of $O(6,22)$ symmetry\cite{GPR} of (\ref{string}).

The class of solutions generated by the $SO(6)$ transformations
on the $U(1)_M \times U(1)_E$ BH solutions corresponds \cite{NONEX}
to charged configurations $\{ P^i, Q^i\}$ ($i=1,...,n(=6)$)
subject to the constraint $\sum_{i=1}^n P^iQ^i=0 $. The most general
solutions within this class, {\it i.e.}, those with unconstrained
charge configurations, are expected to be generated \cite{CY}
by  (one parameter) transformations, belonging to $SO(2,n)$ ($n=6$),
on the former solutions. Here, $SO(2,n)$ is a symmetry of the effective 3-d
Lagrangian density \cite{DM} for the spherically symmetric BH Ansatz
in $(4+n)$-d KK theory.

Such explicit solutions for all static 4-d KK BH's would in turn allow
for a study of their global space-time and thermal properties.
They would in turn provide a sub-class of solutions
for general 57 [or 58]-parameter dyonic [or rotating]
BH solutions which could be generated \cite{SEN} by $[O(22,2)\times
O(6,2)]/[O(22)\times O(6)\times  SO(2)]$ transformations
on the 4-d Schwarzschild [or Kerr] solution.

\acknowledgments
The work is supported by U.S. DOE Grant No. DOE-EY-76-02-3071,
and the NATO collaborative research grant CGR 940870. We would like to thank
C. Kounnas and E. Kiritsis for useful discussions.


\vskip 2.mm

\begin{references}
\bibitem{FIV}{D. Pollard, J. Phys. {\bf A16} (1983) 565; D.J. Gross
and M.J. Perry, Nucl. Phys. {\bf B226} (1983) 29; R.D.
Sorkin, Phys. Rev. Lett. {\bf 51} (1983) 87; G.W.
Gibbons and D.L. Wiltshire, Ann. Phys. {\bf 167} (1986) 201;
Y.M. Cho and D.H. Park, J. Math. Phys. {\bf 31} (1990) 695.}

\bibitem{DM}{D. Maison, Gen. Rel. Grav. {\bf 10} (1979) 717;
P. Dobiasch and D. Maison, Gen. Rel. Grav. {\bf 14} (1982) 231.}

\bibitem{EX}{M. Cveti\v c and D. Youm, UPR-623-T preprint (1994),
hep-th \# 9409119, to be published in Nucl. Phys. B.}

\bibitem{WARSAW}{M. Cveti\v c and D. Youm, UPR-643-T (1994), to appear in the
Proceedings  of the  conference:
{\it From Weak Scale to Planck Scale}, Warsaw, September 22-26, 1994 (World
Scientific 1995, S. Pokorski {\it  et al. eds.}); J. Park, CalTech preprint,
to appear.}

\bibitem{NONEX}{M. Cveti\v c and D. Youm, UPR-645-T preprint (1994),
hep-th \# 9502099.}

\bibitem{SEN}{A. Sen, preprint TIFR-TH-94-47 (1994), hep-th\# 9411187.}

\bibitem{HETE}{J. Maharana and J. Schwarz, Nucl. Phys. {\bf B390} (1993) 3;
J. Schwarz, preprint CALT-68-1815, hep-th \# 9209125);
A. Sen, Nucl. Phys. {\bf B404} (1993) 109.}

\bibitem{SENII}{A. Sen, Int. J. Mod. Phys. {\bf A9} (1994) 3707.}

\bibitem{CHB}{Y.M. Cho, Phys. Rev. {\bf D35} (1987) 2628.}

\bibitem{DUFF}{M.J. Duff, J. Rahmfeld, preprint CTP-TAMU-25-94,
hep-th \# 9406105; see also  M.J. Duff, R.R. Khuri, R. Minasian, J. Rahmfeld,
Nucl. Phys. {\bf B418} (1994) 195.}

\bibitem{GIBB}{G.W. Gibbons and M. J. Perry, Nucl. Phys. {\bf B248}
(1984) 629.}

\bibitem{GPR}{A. Giveon and M. Porrati and E. Rabinovici,
Phys. Rept. {\bf 244} (1994) 77.}

\bibitem{CY}{M. Cveti\v c and D. Youm, work in progress.}

\end{references}
\end{document}